\documentclass{elsart}

\usepackage{natbib}
\usepackage{color}

\usepackage{epsfig}

\usepackage{amssymb}

\begin{document}

\begin{frontmatter}



\title{SPICA infrared coronagraph for the direct observation of exo-planets}


\author{Keigo Enya\corauthref{cor}}, for SPICA working group
\address{
Institute of Space and Astronautical Science, 
Japan Aerospace Exploration Agency, 
3-1-1 Yoshinodai, 
Sagamihara, Kanagawa 229-8510, Japan
}
\corauth[cor]{Corresponding author}
\ead{enya@ir.isas.jaxa.jp}



\begin{abstract}

We present a mid-infrared coronagraph to target the direct observation of extrasolar planets, 
for Space Infrared telescope for Cosmology and Astrophysics (SPICA). 
SPICA is a proposed JAXA-ESA mission, which will carry a telescope cooled to 5K 
with a 3.5m diameter aperture, and is planned to be launched in 2018 by an H\,II
family rocket.
The SPICA mission gives us a unique opportunity for high-contrast observations 
because of the large telescope aperture, the simple pupil shape, 
and the capability for infrared observations from space. 
We have commenced studies for a coronagraph for SPICA,
in which this coronagraph is currently regarded as an option of the focal plane instruments. 
The primary target of the SPICA coronagraph is the direct observation of Jovian exo-planets. 
A strategy of the baseline survey and the specifications 
for the coronagraph instrument for the survey are introduced together.
The main wavelengths and the contrast required for the observations
are 3.5--27$\mu$m, and 10$^{-6}$, respectively.
Laboratory experiments were performed with a visible laser 
to demonstrate the principles of the coronagraphs. 
In an experiment using binary-shaped pupil coronagraphs,
a contrast of 6.7$\times$10$^{-8}$ was achieved, 
as derived from the linear average in the dark region and the core of the 
point spread function (PSF). 
A coronagraph by a binary-shaped pupil mask is 
a baseline solution for SPICA because of its feasibility and robustness.
On the other hand, a laboratory experiment of the phase induced amplitude 
apodization/binary-mask hybrid coronagraph has been executed
to obtain an option of higher performance 
(i.e., smaller inner working angle and higher throughput),
and a contrast of 6.5$\times$10$^{-7}$ was achieved with active wavefront control.
Potentially important by-product of the instrument,
transit monitoring for characteization of 
exo-planets, is also described.
We also present recent progress of technology on a design of a binary-shaped 
pupil mask for the actual pupil of SPICA, PSF subtraction, 
the development of free-standing binary masks, a vacuum chamber, and 
a cryogenic deformable mirror.
Considering SPICA to be an essential platform for coronagraphs
and the progress of key technologies, we propose to develop a mid-infrared coronagraph 
instrument for SPICA and to perform the direct observation of exo-planets with it.

\end{abstract}


\end{frontmatter}

\parindent=0.5 cm


\section{
Introduction
}

\noindent \textit{
The solar planetary system, including the Earth;
} \\ 
\noindent \textit{
Is such a system unique in the Universe, or is it commonplace?
} \\  
\noindent \textit{
How is such a planetary system born?
} \\
\noindent \textit{
And finally, how about the origins of life?
} \\

We consider that these are some of the most important questions 
for space science in the near future.
To answer these questions, considerable efforts have been made. 
Since the first report by \citet{Mayor1995}, 
more than 300 extrasolar planets (exo-planets) have 
been discovered by detailed observations of the Doppler shifts in the spectrum of the 
central star that are caused by the revolution of the planets, 
or from variability in the luminosity of the central star due to the transits 
of these planets 
(\citet{Charbonneau2000}; \citet{Henry2000}). 
Such groud-based observations of exo-planet spatially not resolved from the
central star,
do not usually clarify the spectrum, 
the luminosity and other important properties of the planet itself.
Though recent studies with space telescopes succeeded to derive 
significant spectral features of exo-planets
by observations of the planetary transit
(e.g., \citet{Deming2005}, 
see also Sec.\ref{sec_transit}),
targets of this method are biased for giant hot-Jupiter
close to the central star.
Therefore, a systematic study of spatially resolved observations, 
especially spectroscopy, of exo-planets from their central star is required. 
When such spatially resolved observations of planetary systems 
are attempted, the enormous contrast of luminosity between 
the central star and its planets is a serious problem that needs to be overcome. 
The typical contrast in a mature planetary system is 
$\sim$10$^{-10}$ in visible light, while the contrast is expected to be relaxed
to $\sim$10$^{-6}$  in the mid-infrared \citep{Traub2002}. 
Because of the difficulties that are introduced due to this 
contrast, the detection of exo-planets by spatially 
resolved observations from the central star are limited to only 
a few planets distant from the central star, 
though recent reports concerning such detections are excellently
important milestones in this field 
(\citet{Marois2008}; \citet{Kalas2008}).

The coronagraph, which was first proposed by \citet{Lyot1939}
for the observation of the solar corona, 
is a special optical instrument which we expect could be used to reduce the contrast between 
exo-planets and their central stars. 
For high-contrast observations, the positioning of instruments in space has significant 
advantages over ground-based equipment because the environment in space is 
free from turbulence of the atmosphere, 
which is usually a serious limiting factor of the contrast. 
In recent studies, it is considered that the use of coronagraphs will become one of 
the most promising methods for realizing the observation of exo-planets,
as well as the use of the interferometer or the external occulter
(\citet{Traub2006}; \citet{Beichman2006}; \citet{Cockell2008}; \citet{Cash2006}). 
We consider coronagraphs are the most realistic solution of these three 
for the purpose of the early and efficient realization of the high-contrast 
observation from space, 
if it is designed in the context of being carried as one of instruments 
in a general purpose space-borne telescope for infrared astronomy.

We have focused on the Space Infrared telescope for Cosmology and 
Astrophysics (SPICA) mission which has recently progressed to the pre-project 
phase, which approximately corresponds to Phase-A.
Basic studies have been performed in order to equip SPICA with 
a mid-infrared coronagraph, together with other general-purpose instruments
(\citet{Enya2008a}; \citet{Enya2007a}; \citet{Abe2007}; \citet{Enya2006a}; \citet{Enya2006b}; 
\citet{Abe2006};\citet{Tamura2000}),
as also described in the SPICA mission proposal (\citet{Enya2005}; \citet{Enya2007b}). 
The coronagraph is currently regarded as one option among the focal-plane instruments
\citep{Matsuhara2008}.
As described in Sec.\,\ref{sec_platform},
we consider SPICA has the potential to be a unique 
and essential opportunity for coronagraphic observations in the 2010s.
Studies for critical basic technologies are set to succeed,
as presented in the following sections.
Hence, we propose to develop a mid-infrared coronagraph instrument for 
SPICA and to perform the direct observation of exo-planets using by it. 
This paper summarizes our studies for the SPICA coronagraph, 
accomplished until the kick-off of the pre-project phase of SPICA.

\section{SPICA as a coronagraph platform}\label{sec_platform}

\begin{figure}[!ht]
\begin{center}
\includegraphics*[height=16cm, angle=-90]{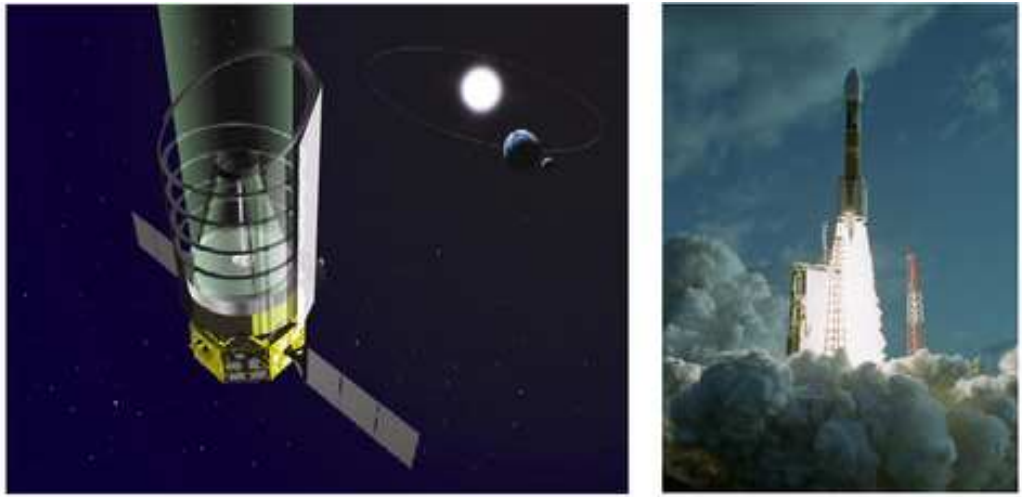}
\end{center}
\caption{
Left: Artists view showing the SPICA telescope in orbit. 
Right: an H\,II-A rocket (\copyright  by JAXA).
}
\label{fig_spica_h2}
\end{figure}

\begin{figure}[!ht]
\begin{center}
\includegraphics*[width=16cm,  angle=180]{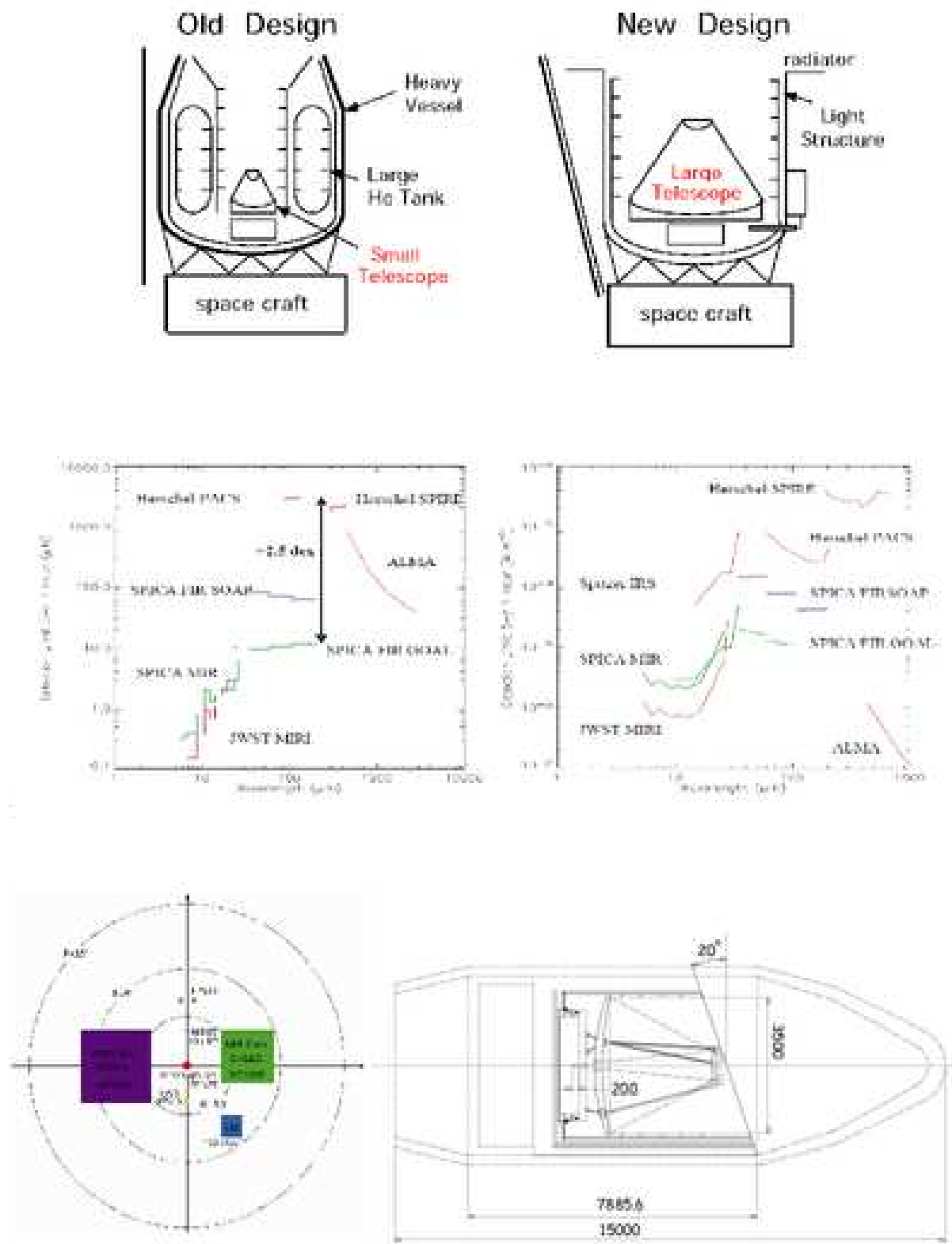}
\end{center}
\caption{
Top: Schematic figures to show the concept of the cryogenics of SPICA. 
Middle: Sensitivity of SPICA telescope for the cases of imaging (left) and spectroscopy (right).
Bottom-left: distribution of focal plane for the expected instrumentation. 
Bottom-right: SPICA in the fairing of an H\,II-A rocket.
}
\label{fig_spica_info}
\end{figure}

SPICA is a space-borne large cooled telescope mission of the next generation
for infrared astronomy and astrophysics,
which was formerly called H\,II/L2 (\citet{Nakagawa2008}; \citet{Nakagawa1998}). 
Table\,\ref{table1} is a summary of the specification of the SPICA mission.
SPICA will carry an on-axis Ritchey-Chretien telescope, which has a monolithic mirror with a 
3.5m aperture as shown in Fig.\,\ref{fig_spica_info}
(\citet{Kaneda2007}; \citet{Onaka2005}). 
The whole telescope is cooled to $\sim$5K by radiation cooling and mechanical cryo-coolers. 
Diffraction-limited optical quality at a wavelength of 5$\mu$m is required for the telescope.
The main observation wavelength is in the range 5--200$\mu$m, and therefore SPICA is basically 
complementary to the James Webb Space Telescope (JWST)
and the Herschel Space Observatory (\citet{Gardner2008}; \citet{Pilbratt2008}), 
while their capabilities are partially overlapped. 
SPICA is planned to be launched into the Sun-Earth L2 libration halo orbit by an H\,II-A 
rocket of JAXA in 2017,  
and will operate for 5 years (Fig.\,\ref{fig_spica_h2}). 
SPICA was proposed to ESA as a Japanese-led JAXA-ESA mission as part of 
Cosmic Vision 2015-2025, and was accepted for an assessment study. 
In the context of JAXA, SPICA has now officially progressed to a new phase, 
the pre-project phase (which approximately corresponds to Phase-A) since July 2008. 
For the Japanese community, SPICA is the third mission for infrared space astronomy, 
following the Infrared Telescope in Space (IRTS)  and AKARI, 
which was launched in February 2006 (\citet{Murakami1996}; \citet{Murakami2007}).

The design concept of the cryogenic system that will be used for 
SPICA is one of the most important reasons why the SPICA telescope can be so large 
as shown in Fig.\,\ref{fig_spica_info} \citep{Sugita2008}. 
SPICA will be warm-launched, and will then be cooled-down in orbit by a combination of radiation 
cooling and mechanical cryo-coolers. 
This approach using no cryogenic liquid is essentially new, 
whereas AKARI carried a liquid helium tank combined with a mechanical 
cryo-cooler which were used 
to cool its 68.5cm telescope to 6.5K (\citet{Kaneda2005}; \citet{Nakagawa2007}), 
and a similar design of cryogenic system is being used 
on the Spitzer Space Telescope, which has an 85 cm aperture \citep{Werner2004}. 
As a result of the elimination of the cryostat and the cryogenic liquid tank, 
it will become possible for the SPICA mission to utilize more of its launch payload and volume 
for the telescope than a conventional infrared space telescope mission. 

The development of a large, light-weight, telescope for cryogenic use is another key 
technology issue to realize the SPICA mission
(\citet{Kaneda2007}; \citet{Onaka2005}). 
The large telescope aperture allows a small diffraction limit (i.e., high spatial resolution)
and high sensitivity, which is essential for the coronagraphic observation.
The on-axis Ritchey-Chretien optics that have been adopted for the SPICA telescope are 
intended to provide a wide field of view, to be simple, and to enable a large (3.5m) 
entrance aperture within the constraints of the dimensions of the H\,II-A rocket fairing. 
Silicon carbide or carbon-fiber reinforced silicon carbide is planned to be used as the 
construction material for the whole of the structure of the telescope, including the primary and
the secondary mirror, the optical bench for the focal-plane instruments and the secondary 
supporting structure. 
Both the primary and the secondary mirrors of the SPICA telescope are monolithic. 
These monolithic mirrors enable the pupil to have a simple shape, 
and a clean point spread function (PSF) free from complicated diffraction patterns, 
expected from
a segmented-mirror telescope.
(It should be noted that the pupil of the SPICA telescope is partly obstructed by the secondary 
mirror and its support structure because it uses on-axis optics, and such an obstruction 
places constraints on the design and performance of the coronagraph, as discussed 
in Sec.\,\ref{sec_spica_pupil}).

It is convenient to summarize here the advantages of SPICA 
for coronagraphic observations; 
1) SPICA is designed for observations in the mid-infrared, 
in which the contrast between the planets and the central star 
is much relaxed than it is at optical wavelengths. 
2) SPICA is unaffected by atmospheric turbulence which is a serious 
limiting factor for the performance of coronagraphs with ground-based
telescopes. 
3) It is also free from the constraints caused by the window 
of atmospheric transmissivity. 
Therefore, continuous spectroscopic coverage of the wavelength 
range of interest is possible. 
4) Clean PSF and high sensitivity are achievable owing to the use 
of a large telescope consisting of monolithic mirrors.

Special space missions have been proposed that target the direct
detection and characterization of terrestrial exo-planets, for example, 
TPF-C \citep{Traub2006}, TPF-I \citep{Beichman2006}, DARWIN \citep{Cockell2008},
and a mission using a large external occulter \citep{Cash2006}. 
These missions are enormous in scale (see also recently proposed smaller scale missions,
e.g., PECO \citep{Guyon2008}), and required technology 
in order to realize ultimate performance is highly challenging.
As a result of various financial, technical and other constraints, 
it seems inevitable that the launches of these huge missions will 
be significantly later than that of SPICA. 
Conversely, JWST is a mission of a general purpose space 
telescope with 6.5m aperture, and its launch is planned  to commence in 2014. 
JWST will carry coronagraphs in two instruments; the Near-Infrared Camera (NIRCAM) 
for 0.6--5$\mu$m observation includes Lyot-type coronagraphs 
\citep{Horner2008}, 
and the Mid-Infrared Instrument (MIRI) for 5--27$\mu$m observation
includes coronagraphs using four quadrant phase masks and a Lyot-type coronagraph
for imaging in bands with 10.65, 11.4, 15.5, 23$\mu$m wavelength 
\citep{Wright2008}. 
Therefore, the wavelength coverage of the coronagraph in JWST overlap with those of SPICA. 
However, the primary mirror of JWST consists of 18 segments, which produces a complicated PSF 
and limits the coronagraphic performance. 
As shown in Fig.\,\ref{fig_sed}, 
contrast is the limiting factor of the detectability of planets rather than 
sensitivity in the short wavelength region (shorter than 10$\sim$15$\mu$m),
in the typical case of the observation by SPICA. 
This short wavelength region is especially important because of the rich spectral features 
that it contains and its small diffraction limit. 
On the other hand, sensitivity limits detectability of exo-planets 
in the long wavelength region (longer than 10$\sim$15$\mu$m). 
The sensitivity of SPICA turns out to be higher than JWST at around 18$\mu$m because of 
the lower temperature of the SPICA telescope (5K) than of JWST (45K). 
As a result, the imaging with the SPICA coronagraph could play a significant role 
in its widely covering wavelength region. 
The capability of the spectroscopy function is the other major unique advantage of 
the SPICA coronagraph. 
Giant ground-based telescopes of the 30m class are planned to start operation in the 2010s.
For example, the first light of the Thirty Meter Telescope (TMT) 
is planned to commence in 2017 \citep{Sanders2008},
and it is expected that the first light of the European Extremely Large telescope (EELT) 
and the Giant Magellan Telescope (GMT) will be realized following the TMT 
(\citet{Gilmozzi2008}; \citet{Johns2008}).
While these giant ground-based telescopes will exceed it 
in terms of spatial resolution with the support of adaptive optics in the next generation,
SPICA still has unique advantages as a platform for a mid-infrared space coronagraph.
In comparison with these forthcoming telescopes, 
we consider that SPICA has the potential to be a unique 
and essential opportunity in the 2010s for the direct observation of exo-planets.





\section{Science cases for the specifications of the coronagraph}

\subsection{The baseline survey and the key specification of the coronagraph}

Table\,\ref{table2} shows the specification of the SPICA coronagraph,
derived as described below. 
In order to determine the specifications of the SPICA coronagraph, 
we need to consider a set of important parameters, its contrast, observation-wavelength, 
inner working angle (IWA), and throughput. 
Whereas extended exposure times do increase the efficiency of throughput,
the relationship between these specific values involves something of a trade off, 
and the requirement for too high a performance leads to unrealistic specifications 
for the entire telescope and the spacecraft. 
Therefore, some estimation of the appropriate requirements is necessary to define these values.

In order to determine these critical specifications for the SPICA coronagraph, 
we considered a baseline survey that was intended as a statistical study of nearby stars by 
coronagraphic observation. 
In comparison with the coronagraphs that will be fitted to the JWST and to large ground-based 
telescopes, the primary advantage of the SPICA coronagraph is its higher contrast 
in the mid-infrared wavelength region, while its spatial resolution is lower. 
So the SPICA coronagraph has the potential to perform unique work in high-contrast surveys 
of nearby stars for the direct observation of not only warm but relatively 
cooler Jovian exo-planets. 
Such deep coronagraphic 
observations are useful when we are trying to reveal common exo-planetary systems, 
because the aim is not only to find special big and/or young exo-planets, 
but also to detect older objects. 
Statistical studies are required to understand the common nature of exo-planetary systems,
which will be more important especially if the first clear direct detection of an exo-planet 
is achieved by either JWST or by a large ground-based telescope before the launch of SPICA.
Statistical analysis is also essential to determine significant constraints 
on the properties of exo-planetary systems even from negative observational results 
(e.g., it may be possible to assign limits on the production rate, 
the frequency distribution of mass, the radius and shape of their orbits, 
and their cooling time-scales). 
On the other hand, we have to consider that the specification of the coronagraph is also 
constrained by the physics of each coronagraphic method, which should therefore be selected 
based on their expected performance, feasibility, and robustness for actual use in SPICA. 
Considering these factors, the critical parameters of the coronagraph, the baseline survey, 
and the coronagraphic method were derived together as shown below.

\begin{figure}[!ht]
\begin{center}
\includegraphics*[width=8cm]{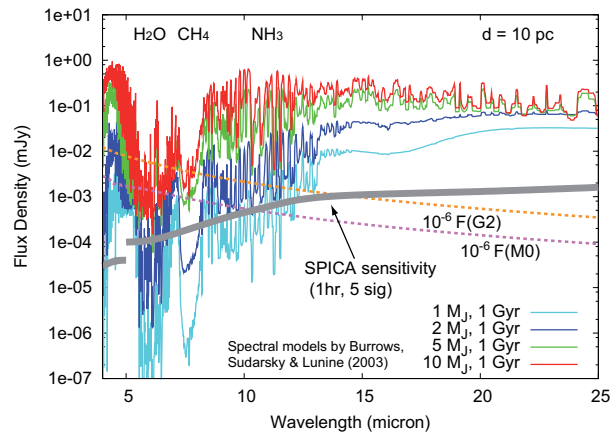}
\end{center}
\caption{
Calculated SED of 1Gyrs-old Jovian planets with various mass 
presented as the Fig.\,13 in \citet{Burrows1997}, 
and properties relating to SPICA observations. 
10pc is assumed as the distance to the planetary system. 
The light blue, deep blue, and gray curves show the SED of the Sun 
at 10pc scaled down by 10$^{-6}$,  10$^{-7}$,  
and 10$^{-8}$, respectively.
The red curve shows the sensitivity limit of imaging with SPICA. 
The green and purple arrows show the wavelength coverage of a Si:As 
and of a InSb detector, respectively. 
While a Si:As detector has sensitivity at wavelengths even 
shorter than 5$\mu$m, 
an InSb detector reinforces the sensitivity in this wavelength region.
Figure is from \citet{Fukagawa2009}.
}
\label{fig_sed}
\end{figure}

\begin{itemize}
\item
Observation mode:\\
The use of imaging mode, rather than the spectroscopy mode, 
is presumed for the baseline survey.
\\

\item
Number of targets and distance to them:\\
For statistical studies of exo-planets, we require information 
about $\sim$100 nearby stars as the targets for a baseline survey. 
This means that the observation of stars at around $\sim$10pc working distance is required. 
If we consider the probability of a star having planets and the fact that the position 
of a planet in its orbit is not always suitable for observation, a few 10s of stars not enough. 
On the other hand, 1000 or more targets will be too many for a non-biased survey 
in practical terms.
\\

\item
Observation wavelength:\\
3.5--27$\mu$m is supposed. 
Fig.\,\ref{fig_sed}
shows a simulated spectrum of Jovian exo-planets in the mid-infrared \citep{Burrows1997}. 
Rich spectral features are distributed in the short wavelength region; $\sim$3.5--15$\mu$m,
which is also important because of the small diffraction limit. 
A Si:As detector is required to cover this short wavelength region. 
The longest wavelength of the observation wavelengths (27$\mu$m) is determined by the sensitivity 
limit of the Si:As detector. 
As a result, 3.5--27$\mu$m is determined as the total range. 
It should be noted that the SPICA telescope itself is specified for 5--200$\mu$m, 
so the coronagraph instrument is intended to sense infrared in the wavelength range 
shorter than 5$\mu$m by its own, possibly supported by wavefront correction. 
The use of an InSb detector has also been considered as an option to improve the system 
sensitivity in the spectral region between 3.5--5$\mu$m, where the efficiency of Si:As 
detectors is not very high. 
If we adopt the InSb detector option, sensitivity at even shorter wavelengths 
(down to 1$\mu$m) can be acquired.
To realize this option, HgCdTe detectors have also exhibited 
suitable sensitivity; for example, as used in JWST/NIRCAM and NIRSPEC. 
However, we consider that the use of an InSb detector is the best 
current solution for this option because of our experiences 
with using them for cryogenic operations. 
The operating temperature of the detector should preferably 
be as low as possible, down to the temperature of the telescope itself. 
In AKARI, we operated an InSb detector at 10K in space \citep{onaka2007}.

\item
Contrast:\\
The requirement for contrast is $10^{-6}$ in the wavelength region between 3.5--27$\mu$m,
which is defined as the ratio of the intensities between the core and the brightest 
position in the dark region of the PSF. 
The application of the PSF subtraction method can improve the final contrast; for example, 
the improvement was a factor of $\sim$10 in the case of the laboratory experiment shown 
in Sec.\,\ref{sec_subtraction}.  
Such a high-contrast is essential on the shorter side of the wavelength region in the 3.5--27$\mu$m
range in order to detect old ($\sim$Gyrs old) planets (see Fig.\,\ref{fig_sed}). 
On the longer side of the wavelength region between 3.5--27$\mu$m, 
the sensitivity limit of the SPICA telescope, rather than the contrast,  
dominates the detectability of planets.
\\

\item
Coronagraphic method:\\
A coronagraph by a binary shaped pupil mask is the current baseline design 
because of its physical properties and feasibility.
We are also evaluating a Phase Induced Amplitude Apodization PIAA/binary-shaped pupil mask 
hybrid (PIAA/binary-mask hybrid) coronagraph 
as a high performance option to obtain smaller IWA and higher throughput. 
Detail of the study for the selection of these coronagraph methods is 
shown in Sec.\,\ref{sec_exp}. 
\\

\item
IWA:\\
This parameter is strongly limited by the physics of each of the coronagraphic methods. 
In the case of a binary pupil mask coronagraph, IWA$=3.3\lambda/D$.
This limit corresponds to the radius of the orbit of exo-Jupiter at $\sim$5pc distance or 
exo-Saturn at $\sim$10pc distance observed at the wavelength of 5$\mu$m. 
Using the PIAA/binary-mask hybrid option, 
it is expected in principle that IWA will be reduced to $\sim$1.5$\lambda/D$.
\\

\item
Throughput:\\
In this paper, we define the ratio of the input and output fluxes of the coronagraph 
as the throughput. 
Vignetting due to some obstruction of the pupil of the telescope by the secondary mirror 
and its supports are not included in the coronagraph throughput (i.e., the throughput is 
100\% if the instrument consists of only an ideal imager without any masks). 
The throughput of the coronagraph is limited by adopting a particular coronagraph method. 
The specification for the throughput is $\sim 20$\% and $\sim 80$\%
in the case of coronagraphs 
constructed using a binary pupil mask and a PIAA/binary-mask hybrid, respectively. 
\\

\end{itemize}

It is a natural extension of the baseline survey in imaging mode,
to execute intensive follow-ups which adopt spectroscopy-mode for the exo-planets 
and other objects detected in the baseline survey. 
The detection of spectral features relating to H$_2$O (around 6$\mu$m),  
CH$_4$ ($\sim$7.7$\mu$m), 
O$_3$ ($\sim$9.6$\mu$m), NH$_3$ ($\sim$10.7$\mu$m),  CO$_2$ ($\sim$15$\mu$m),
and other features has been tested for the bright exo-planets. 
The bump in the spectral energy distribution (SED) at 4--5$\mu$m wavelength is 
an important feature (\citet{Burrows2004}; \citet{Burrows1997}), 
which is considered to be caused by the opacity window of the atmosphere of the 
planet (i.e., looking at inner and warmer regions). 
Once such a spectral feature is obtained, it should be compared with
those of the other planets in the solar system, 
and the exo-planets known in advance
(e.g., spectral feature of H$_2$O \citet{Tinetti2007}, CH$_4$
\citet{Swain2008a} on HD\,189733b obtained by the transit monitor method. 
See also Sec.\ref{sec_transit}).
If we can obtain various spectra for a significant number of exo-planets
then a correlation 
study will be possible, in which the spectral features of the planets can be related
with their other properties and/or the central stars 
(e.g., mass, age, orbit, existence of inner planets found by the
observations of the Doppler shifts or planetary transit, and so on). 
Such data has the potential to enable the acquisition of an essential 
understanding of the properties and the nature of the evolution of 
planetary systems.

For these spectroscopic observations,
a spectral resolution of
100--200 is needed to detect the spectral features of bright exo-planets,
as shown in Fig.\,\ref{fig_sed}. 
Because of the constraints of sensitivity, it is necessary to adopt an 
appropriate (not too large) value for the spectral resolution in order
to realize efficient observations.

\subsection{Other cornagraphic observations}

\begin{itemize}

\item
Detailed imaging and spectroscopy of the nearest stars:\\
The observation of the nearest stars can be a target of special interest, 
though such observations are included in principle in the baseline survey. 
For example, in the case of $\alpha$ Centauri which is 1.34 pc distant from the Sun, 
the value of IWA for the binary mask coronagraph is close to 1\,AU\,(!) in the 3.5--5$\mu$m
wavelength region. 
Though the observation of objects very close (or marginal) to the IWA is very challenging, 
and although other stars are much further away than $\alpha$ Centauri, 
such observations should be considered well to detect rocky planets at a distance 
of $\sim$1AU from the nearest star. 
As shown below, 
the PIAA/binary-mask hybrid option, PSF subtraction, 
and the InSb detector options can 
be helpful in this quest.
\\

\item
Imaging and spectroscopy of targets in a star-forming region:\\
In a young star-forming region, it will be reasonable to expect that any 
planets will also be young and will be noticeably brighter in the infrared than 
old exo-planets of $\sim$Gyrs old. 
However, typical star forming regions are beyond the coverage distance of the baseline survey, 
which is limited to 10pc. 
For example, the Taurus molecular cloud including class I, II, III stars
is $\sim$140 pc distant from the Sun,
in which young ($\sim$ order of Myrs old) planetary systems can be target \citep{Kenyon1994}. 
In observation of such targets, the detectable planets are constrained to only the outer planets 
due to the limits imposed by the IWA, which corresponds to $\sim$100AU in the 3.5--5$\mu$m
wavelength region if a binary mask coronagraph is used to target the Taurus molecular cloud.
\\

\item
Imaging and spectroscopy of ``known'' exo-planetary systems:\\
More than 300 exo-planets have been found by the radial velocity and/or 
the transit methods (\citet{Mayor1995}; \citet{Henry2000}; \citet{Charbonneau2000}). 
These exo-planetary systems that are already known are potentially interesting targets. 
Furthermore, any new exo-planetary systems that might be found prior 
to observation with SPICA are also important targets. 
COROT, which was launched in December 2006 \citep{Barge2008}, 
and Kepler, which has been launched in March 2009 \citep{Borucki2008}, 
are satellites to observe the transit of exo-planets. 
Some systematic observations by direct imaging with large ground-based telescopes 
are ongoing, for example, NICI/VLT \citep{Toomey2003}, 
HiCHIAO/SUBARU (\citet{Tamura2006}; \citet{Hodapp2006}). 
JWST is a powerful tool in the search for exo-planets 
before the observation by SPICA. 
For all of these ``known'' exo-planetary systems, 
it can be essential for SPICA to attempt to implement direct imaging/spectroscopy.
Most of planets detected by the radial velocity or 
the transit methods are expected to be biased towards being 
close to the central star, and so direct detection of them is difficult. 
Conversely, observations with the SPICA coronagraph are complementary to these methods, 
and therefore SPICA will make it possible to study the correlation between properties of the known
planetary systems and new information derived by SPICA,
e.g., previously undetected planets in these systems
at grater distances from the star. 
For exo-planets detected directly before SPICA, follow-up spectroscopy 
with the mid-infrared coronagraph of SPICA can be useful to characterize 
the planets.
\\

\item
Imaging and spectroscopy of disks and exo-zodiacal light:\\
Proto-planetary disks and debris disks are robust targets for the SPICA coronagraph. 
As well as those disks that are already known, any disks that might be 
discovered before SPICA will also become important targets. 
We expect a baseline survey with SPICA will newly find many disks.
Disks that are detected by ground-based telescopes and/or by 
JWST will all be potential targets. 
Imaging/spectroscopy with higher contrast in the mid-infrared by the SPICA coronagraph 
will be a powerful tool, especially for the detection and characterization of 
faint disks  and of exo-zodiacal light that are not detected in studies
of SED excesses, and for the study of spatial structure 
(i.e., rings, voids) in the outer disks in the mid-infrared
as signpost of planet formation and exsistence.
\\

\item
Other targets:\\
Not only exo-planetary systems and disks, but any other interesting objects with 
high-contrast can be targets, for example, brown dwarfs, host galaxies of active 
galactic nuclei and quasars.
\\

\item
Monitor observations:\\
Many of the targets listed above are potentially variable. 
Movements of the planets in their orbits due to revolution and seasonal variations may 
be observable in the timescale of the SPICA mission. 
The timescale of the rotation of a planet on its axis is expected to be shorter and 
to be more suitable for monitoring observations. 
It will be very exciting if such periodic and/or irregular variability in position, 
luminosity, and spectral features can be confirmed on the properties of exo-planets.
\\

\end{itemize}

The performance of these observations can be enhanced by various optional 
functions and their operations. 
The PIAA/binary-mask hybrid option reduces the IWA down to $\sim$1.5$\lambda/D$ and increases 
the throughput up to $\sim$80\%.
If an InSb detector is applied, then sensitivity in the 3.5--5$\mu$m range will be improved, 
sensing at shorter wavelengths down to $\sim$1$\mu$m becomes possible, 
and smaller values for IWA will be obtained via decreasing values of $\lambda/D$.

On the other hand, the advantages in contrast in the mid-infrared are 
progressively lost in the shorter wavelength regions. 
Wavefront errors (WFE) and tip-tilt errors are more serious owing to the effects 
of the factor of $\lambda/D$ in the shorter wavelength regions, and therefore the PSF contrast 
is expected to be worse. 
PSF subtraction is essential in these observations to cover the disadvantages 
in terms of contrast. 
If the stability of the system is very good, 
as can be expected in a thermally stable,
cryogenic environment in space,
then the IWA and the sensitivity after 
the PSF subtraction are potentially going to become useful in those regions where 
rocky planets may exist at 1AU distance from the nearest stars.

\subsection{Monitoring of planetary transit}\label{sec_transit}

One of the most interesting by-products of this instrument is the
monitoring of transiting exo-planets which it targets to characterize
the spectral features. 
Both the primary and the secondary eclipses can be targeted. 
The spectrum of a transiting planet is obtained by the
spectral difference between ``in'' and ``out'' of transit. 
In principle, observation of planetary transit can provide the spectrum
of an exo-planet which is too close to its central star to allow
detection by direct observation. 
Conversely, the frequency of transit decreases rapidly 
if the radius of orbit of the planet becomes large. 
Therefore coronagraphic direct observation and transit
monitoring are complementary.

Space-borne telescopes have pioneered the characterization of the
spectra of exo-planets by the transit monitoring method. 
Spitzer Space Telescope revealed the SED of 
HD209458b \citep{Deming2005}, 
TrES-1 \citep{Charbonneau2005},
HD189733b  \citep{Deming2006}, 
$\upsilon$ Andromeda b \citep{Harrington2006}. 
This telescope was also used to obtain spectrum of 
HD 209458b (\citet{Richardson2007}; \citet{Swain2008b}) and 
HD189733b (\citet{Grillmair2007}; \citet{Tinetti2007}).
The Hubble Space Telescope was also used to study the spectrum 
of HD189733b \citep{Swain2008a}.
Studies and discussions about the spectrum, 
including important features, the H$_2$O and CH$_4$ lines, 
are ongoing (\citet{Desert2009}; \citet{Sing2009}). 
These quite interesting results show that this method is one 
of the most promising ways for enabling the characterization of 
exo-planets.

SPICA, with a 3.5m aperture, will gather many more photons than the
Spitzer Space Telescope, with an 85cm aperture. 
This advantage drastically improves sensitivity in the case 
in which the residual spectrum, after subtraction, is
photon-noise-limited. 
Non-coronagraphic observation for transit monitoring will be realized 
by a mechanical mask changer or the static forking of the optical 
path in the instrument, enabling the potential
to add the capability of a fine pixel-scale imager and a spectrometer
to the focal plane instruments of SPICA. 
As shown in Table\,\ref{table2}, imaging or spectroscopy with spectral resolution 
of 20--200 at 3.5--27$\mu$m wavelength is available for planetary 
transit monitoring. 
If the InSb detector option is adopted, simultaneous coverage of 
wavelength is extended down to $\sim$1$\mu$m.

Targets having planetary transits should be found from special surveys
for transit detection. 
COROT, which is presently working in orbit \citep{Barge2008}, 
and Kepler, which was launched in 2009
\citep{Borucki2008}, are satellites just for transit survey. 
It's expected that these missions will provide a sufficient number of 
targets (more than $\sim$1000 stars with planetary transit) 
before a detailed characterization with SPICA.

It should be noted that telescope aperture of JWST (6.5m) is larger
and the planned launch of JWST (in 2014), is earlier than SPICA. 
One of the potential advantages of SPICA is stability, 
which is essential for bright targets in which systematic error 
dominates over photon-noise. 
In the SPCIA telescope system, the monolithic mirrors contribute 
to optical stability. 
The SPICA telescope is surrounded by baffles and a Sun-shield, 
which contribute to temperature stability. 
Whole of the telescope, made of the SiC family of materials, 
will be cooled to $\sim$5K in the SPICA mission. 
The coefficient of thermal expansion of the 
telescope material is very small at this temperature \citep{Enya2007d},
which provides stability against thermal deformation in observation. 
The study of the instrument design, intended for use in
transit observation, has been started. 
Deep full-well of the detector and rapid readout both 
improve the duty cycle of exposure. 
An internal calibration source is expected to be quite useful 
for calibrating the instability in the sensitivity of the 
detector array during monitoring observations. 
Simultaneous wavelength coverage in $\sim$1--27$\mu$m
wavelength range will also be of advantage for SPICA for monitoring
planetary transits if a InSb detector option is adopted as the 
secondary detector for the short wavelength channel together
with a Si:As detector.
JWST needs NIRCAM\citep{Horner2008} and MIRI\citep{Wright2008} 
to cover this wavelength region,
however, these two instruments do not have common field of view
and therefore simultaneous use for a point source is not possible.
In the focal plane instruments of SPICA, 
the coronagraph instrument is unique for the monitoring observation 
of planetary transit.
InSb detectors which extends the shortest wavelength range of the 
mid-infrared observation down to $\sim$1$\mu$m will not be adopted for
other mid-infrared instruments \citep{Matsuhara2008}.
Furthermore, the telescope pointing stability on a image took 
by the coronagraph instrument is best ($\sim$0.03 arcsecond)
owing to the tip-tilt mirror system,
which is advantage to realize high accuracy photometric monitoring 
over ununiformity of sensitivity in the detector pixels.

\section{Development of coronagraph}
\label{sec_exp}

\subsection{Selection of coronagraphic method}

At the beginning of the study for the SPICA coronagraph, 
many coronagraphic techniques had already been presented
(e.g., summary in \citet{Guyon2006}). 
Such coronagraphic methods were mainly intended for the direct observation of terrestrial 
exo-planets, and it was shown theoretically that many of these methods could provide 
contrasts of 10$^{-10}$. 
However, experimental demonstrations of these systems were poor. 
Thus, we considered that laboratory demonstrations of its capabilities would be one of 
the most important issues for the SPICA coronagraph, even though the targeted contrast was 
``only'' 10$^{-6}$.
When selecting the coronagraphic method that would be used for SPICA, 
there were some particular requirements to consider. 
The SPICA coronagraph had to work in the mid-infrared wavelength region in a cryogenic 
environment, so a coronagraph without transmissive devices was thought to be preferable. 
The coronagraph should be robust against any telescope pointing errors caused by 
vibration of the mechanical cryo-cooler system and the altitude control system of the satellite. 
Achromatism was also assumed to be a beneficial property.

After those considerations, a coronagraph implemented by a binary-shaped pupil mask was 
selected as the primary candidate to be studied because of its physical properties and its 
feasibility 
(\citet{Jacquinot1964}; \citet{Spergel2001}; \citet{Vanderbei2003}; \citet{Vanderbei2004};
\citet{Green2004};
\citet{Kasdin2005a}; \citet{Kasdin2005b}; \citet{Belikov2007}; 
\citet{Tanaka2006a}; \citet{Tanaka2006b}).
We performed laboratory demonstrations with high precision binary 
pupil masks. 
On the other hand, a study of a coronagraph implemented by PIAA using two mirrors presented in 
\citet{Guyon2003}, \citet{Traub2003}, 
was also initiated in order to obtain higher performance
(i.e., smaller IWA and higher throughput).
We introduced a PIAA/binary-mask hybrid solution, 
and it was tested in a laboratory developed by O. Guyon and his colleagues.
Now we consider the PIAA/binary-mask hybrid solution as a hopeful candidate
of the option to realize higher performance over the binary mask mode.
A Prolate Apodized Lyot Coronagraph (PALC) has also been studied, 
which in principle is similar in terms of feasibility to the classical (i.e., well-studied) 
Lyot coronagraph and has the potential to realize smaller IWA, $\leq$2$\lambda/D$
(\citet{Aime2002}; \citet{Soummer2003}; \citet{Aime2004}). 
We have manufactured apodizer masks and have demonstrated 
a Multiple Stage PALC (MS-PALC), in which the second stage significantly improved the performance. 
More detail about each of the experiments is shown below.

It should be noted that some transmissive devices 
(e.g., glass substrates and lenses) were used for convenience 
in the first trials of our laboratory experiments 
because the purpose of the experiment is to demonstrate the principles 
of the coronagraph. 
The development of coronagraphs without transmissive devices 
followed later, as shown in Sec.\ref{sec_mirc}.

\subsection{ 
Laboratory experiments with a binary-shaped pupil mask coronagraph
}\label{exp_ckb}

\begin{figure}[!ht]
\begin{center}
\includegraphics*[width=15cm, angle=180]{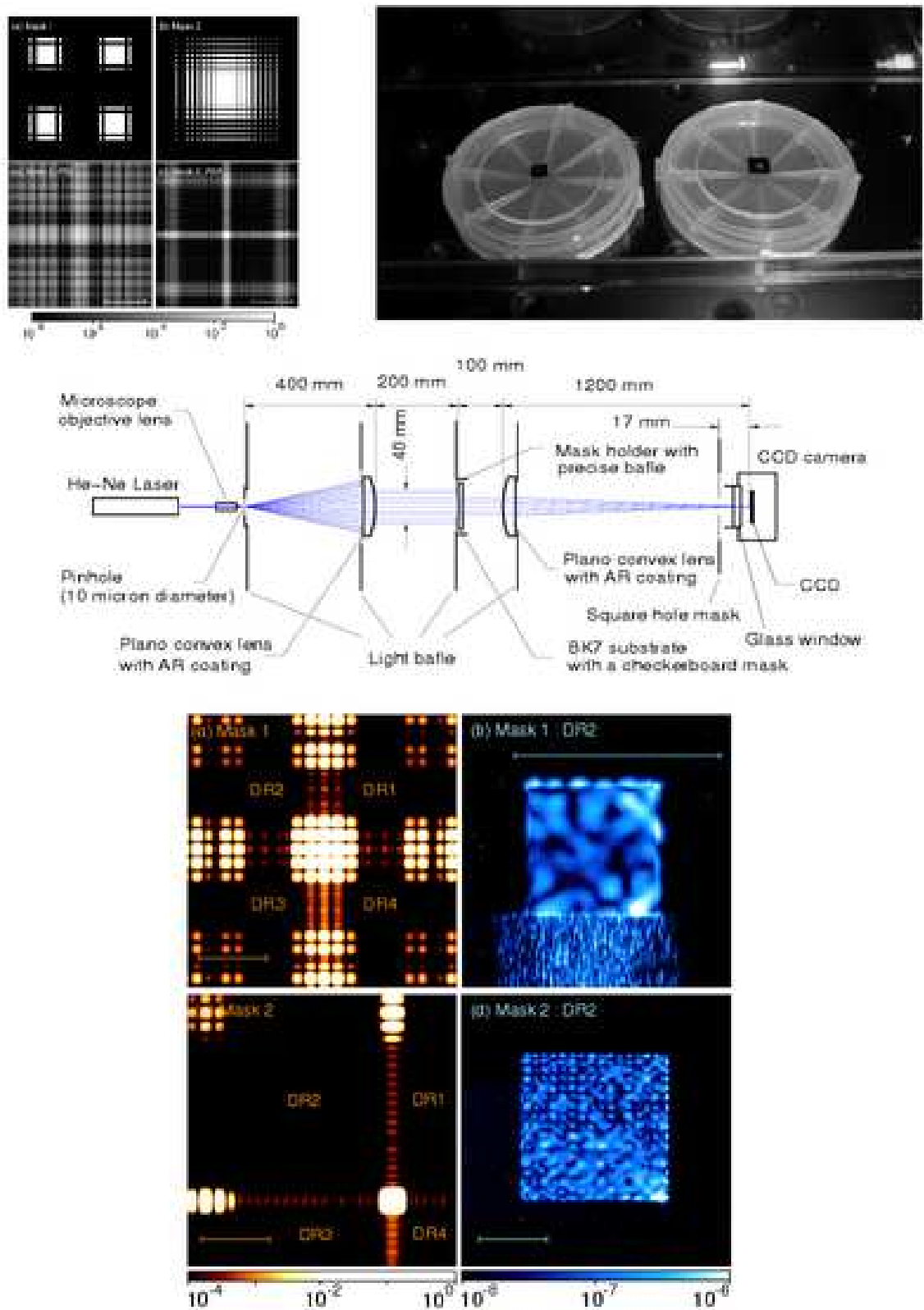}
\end{center}
\caption{
Figures showing experiment with checkerboard Mask-1 and Mask-2. Top-left: designs of Mask-1 
and Mask-2 (top), and expected PSF derived by simulation (bottom). Top-right: manufactured 
masks on BK7 glass substrates. Middle: optical configuration of the experiment. 
Bottom: observed coronagraphic images. Panels (a) and (c) show images including the core of the PSF. 
Panels (b) and (d) are images of the dark region obtained with a mask with a square aperture. 
The scale bar shows 10$\lambda/D$.
}
\label{fig_ckb_exp}
\end{figure}

We carried out experiments to investigate and demonstrate the performance 
of a coronagraph using a checkerboard mask with and without a large central obstruction
(\citet{Vanderbei2004}; \citet{Tanaka2006a}; \citet{Tanaka2006b}), 
which was conducted in visible light in an atmospheric ambient.
Optimization of all of our checkerboard masks was performed using the LOQO solver
presented by \citet{Vanderbei1999}.
Two binary masks, Mask-1 and Mask-2, consisting of 100nm thick aluminum 
patterns on BK7 glass substrates were made by nano-fabrication technology using electron 
beam patterning and a lift-off process (Fig.\,\ref{fig_ckb_exp}). 
These masks were fabricated in collaboration 
with the Nanotechnology Research Institute of Advanced Industrial Science and Technology (AIST). 
Mask-1 is a symmetric checkerboard-type with a central obstruction corresponding to the 
secondary mirror, which has a diameter equal to 30\% of that of the primary mirror. 
Mask-2 is optimized for the case without the central obstruction, 
while the other specifications of Mask-2 in fabrication are the same as those of Mask-1. 
The required contrasts in the design of Mask-1 and Mask-2 were 10$^{-7}$, 
since the target contrast of the demonstration was 10$^{-6}$, 
with a factor of 10 as a margin. 
The values of IWA and outer working angle (OWA) for this design were 7 and 16$\lambda/D$ for Mask-1, 
and 3 and 30$\lambda/D$
for Mask-2, respectively. An anti-reflection (AR) coating was applied to both sides of the masks. 
The configuration of the experiment is shown in Fig.\,\ref{fig_ckb_exp}. 
The experiment was set-up in a 
dark-room with an air-cleaning system employing HEPA filters. The optical devices were 
located on a table fitted with air suspension. A He-Ne laser with a wavelength of 632.8nm 
was used as a light source, with a spatial filter consisting of a microscope objective 
lens and a 10$\mu$m diameter pinhole. 
The laser beam was collimated by a plano-convex lens with an AR coating and the beam was 
irradiated onto the checkerboard mask. After the beam passed through the mask, 
it was focused by a lens with the same specification as the collimating lens but with a longer focal length, 
as shown in Fig.\,\ref{fig_ckb_exp}. 
A cooled CCD camera was used to obtain the images. Wavefront control by adaptive optics 
was not applied. 
We used combinations of several different exposure times and optical density filters 
to expand the dynamic range of the measurements. 
For the measurements of the dark region of the PSF, a black-coated mask of SUS304 which 
was 50$\mu$m in thickness and with a square hole was set in front of the CCD camera to 
block-out the flux of the PSF core to reduce the scattered light in the camera. 
The bottom panel of Fig.\,\ref{fig_ckb_exp}
shows the observed PSF with the checkerboard mask coronagraph. 
For both Mask-1 and Mask-2, the shape of the core of the PSF closely resembles the designed shape. 
The PSF contrasts,  
derived from the linear-scale average measured in the dark region and the core of the PSF,
are 2.7$\times$10$^{-7}$ for Mask-1, and 1.1$\times$10$^{-7}$ for Mask-2, respectively. 
Both of the IWA and OWA values that were obtained from 
the experiments are also consistent with the design values. Therefore, it was concluded that 
a coronagraph with a binary checkerboard pupil mask with or without a large central obstruction 
works in terms of our target contrast (10$^{-6}$). We also concluded that speckle is the primary 
limiting factor which affects the coronagraphic performance in this experiment, 
as discussed in \citet{Enya2007c}.

\subsection{
Laboratory experiment of PIAA/binary-mask hybrid coronagraph
}

\begin{figure}[!ht]
\begin{center}
\includegraphics*[width=16cm, angle=180]{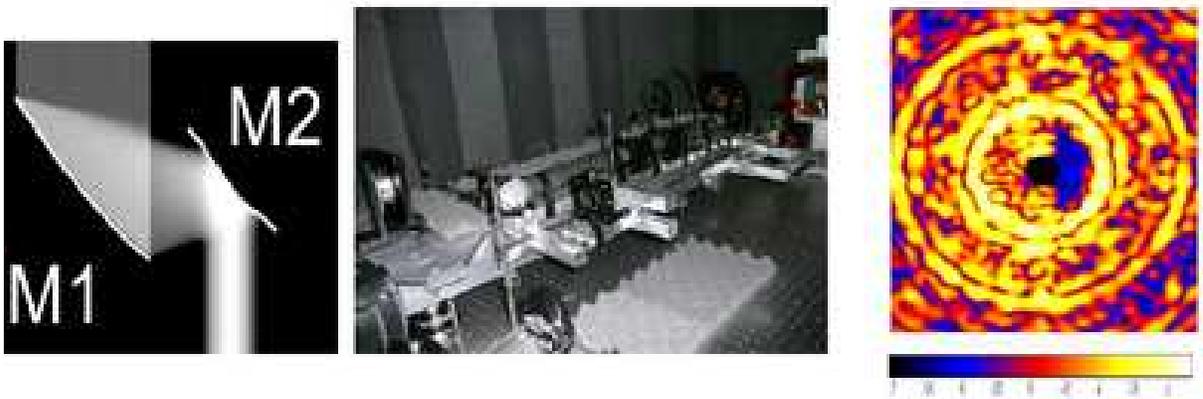}
\end{center}
\caption{
Left: schematic figure to show principle of PIAA. 
Middle: optical configuration used for the laboratory demonstration 
of a PIAA/binary-mask hybrid coronagraph. 
Right: observed coronagraphic image shown on a log scale. 
It must be noted that these results were obtained at a laboratory 
pioneered by O. Guyon and his colleagues at the SUBARU observatory
(e.g., \citet{Tanaka2007}). 
The core of the PSF is obstructed by a focal plane mask 
with a radius of 1.5$\lambda/D$
A dark region is produced to the right of the core by implementing 
wavefront control using a deformable mirror with 32$\times$32 actuators. 
The average contrast of the dark region is 6.5$\times$10$^{-7}$.
}
\label{fig_piaa}
\end{figure}

More recently, experiments involving a coronagraph employing PIAA have also been progressed. 
The PIAA coronagraph uses two mirrors to realize apodization for high-contrast imaging needed 
for use in exo-planet searches (\citet{Guyon2003}; \citet{Traub2003}). 
In principle, it achieves a very high throughput ($\sim$100\%) 
and a quite small IWA ($\leq$1.5$\lambda/D$) simultaneously. 
However, a PIAA that was designed to provide high-contrast by itself would suffer from 
optical shapes that are difficult to polish, as well as reduced bandwidth due to chromatic 
diffraction. 
Both of these problems can be solved simultaneously by adopting a hybrid PIAA 
design, 
in which the apodization created by the two-mirror system is moderated by combining 
it with a classical apodizer with PIAA apodization \citep{Pluzhnik2006}. 
\citet{Tanaka2007} presented an implementation of a hybrid PIAA system 
with binary-shaped masks
of concentric ring type, and showed in the laboratory that such a combination 
is a robust approach to high-contrast imaging.
The laboratory experiment was performed in air using a He-Ne 
laser as shown in Fig.\,\ref{fig_piaa}. 
The entire optics for the experiment was set-up in a clean-room
and all the experiment was carried out in the air,
which are same style to the experiments with a binary-shaped pupil mask coronagraph.
Active wavefront control was applied using a deformable mirror  
with 32$\times$32 channels in Micro Electro Mechanical Systems (MEMS) technology, 
which was commercially provided by Boston Micromachines Corporation (BMC). 
As shown in Sec.\ref{exp_ckb}, neither air turbulence nor instability of 
optics caused by temperature fluctuations are seen as problems 
that would prevent us from demonstrating our goal contrast of  10$^{-6}$. 
The quality of the light source is also regarded as being adequate 
to demonstrate a contrast of 10$^{-6}$. 
Therefore, the role of wave-front control in this experiment 
is to cancel out any imperfections in the PIAA optics, 
i.e., the surface figure of the primary and the secondary 
mirrors of the PIAA, the combined binary-shaped mask, 
and their alignment.

The contrast reached 6.5$\times$10$^{-7}$ at a separation of $\sim$1.5$\lambda/D$  
from the center of the PSF (Fig.\,\ref{fig_piaa}). 
It must be highlighted that the results of this experiment were obtained at a 
laboratory pioneered by O. Guyon and his colleagues at the SUBARU observatory 
\citep{Tanaka2007}.

\subsection{
Laboratory experiment of MS-PALC
}

MS-PALC optics could be realized by the combination of sets of an occulting mask at 
the focal plane with a Lyot stop as a classical Lyot coronagraph, and a prolate apodized pupil
(\citet{Aime2002}; \citet{Soummer2003}; \citet{Aime2004}). 
It is based on a property of the prolate apodized function, whereby the result of 
transformation through a single stage of PALC, including
the Fourier transformation and masking maintains the properties of the prolate function. 
This fact means that the total contrast of the MS-PALC optics becomes higher as a function 
of the number of stages. 
In principle, MS-PALC has similar feasibility to that of a classical (i.e., well studied) 
Lyot coronagraph, and also has the potential to realize a smaller 
IWA ($\leq$2$\lambda/D$ by two stages) 
with a very small occulting mask at the focal plane, of which radius 
is smaller than 1$\lambda/D$. 
Thinking about these advantages, 
we started an experimental study of the MS-PALC in the context of work related 
to the SPICA coronagraph \citep{Abe2008}. 
We adopted the use of High energy beam sensitive (HEBS) glasse to 
realize a prolate apodization mask, 
which was provided by Canyon Materials Incorporated (CMI). 
A layer in the HEBS glass, which includes Ag$^+$ ions, 
is sensitive to the electron beam, and is specified such that 
it is patterned with 500 gray levels with $\sim$0.1$\mu$m spatial 
scale thanks to the resolution of the electron beam 
(see also the website of CMI to see more detail about HEBS glass).

\begin{figure}[!ht]
\begin{center}
\includegraphics*[width=14cm, angle=180]{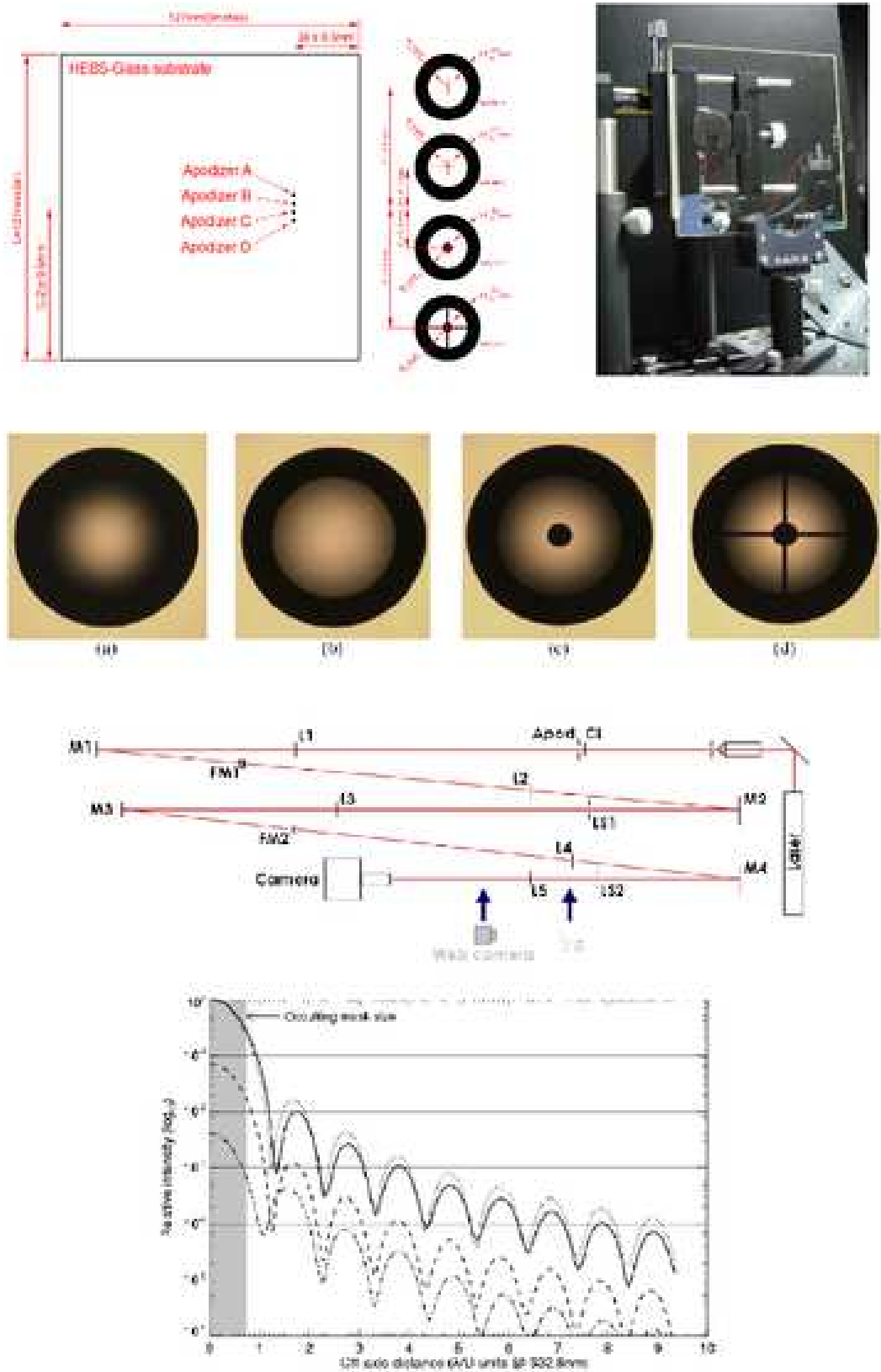}
\end{center}
\caption{
Figures showing experiment of MS-PALC. Top: design of the HEBS glass apodizer (left panel) 
and manufactured apodizer in optics. Upper-middle: microscope photographs of the 
apodizers taken with transmitted light. Upper-middle: optical configuration of the experiment. 
Bottom: Profiles of the coronagraphic images obtained by the MS-PALC experiment.
}
\label{fig_palc}
\end{figure}

We designed 4 apodizers, as shown in Fig.\,\ref{fig_palc}. 
Apodizer-A and -B have no obstructions, 
while Apodizer-C and -D have a central obstruction and 
furthermore Apodizer-D has an obstruction in the form of a spider pattern. 
2 mm diameter versions of Apodizer-A, -B, -C, and -D were fabricated on 
5\,inch$\times$5\,inch substrates made of HEBS glass. 
The configuration of the optics is shown in Fig.\,\ref{fig_palc}. 
A He-Ne laser with a wavelength of 632.8nm was used as the light source for the experiment. 
Lenses made of BK7 were used to collimate and focus the beam, and the long light-path 
was compacted by mirrors. 
Circular focal-plane masks made of aluminum were manufactured on BK7 substrates. 
Intensive evaluations were performed in the case of Apodizer-B using focal plane masks with 
radii corresponding to 0.75$\lambda/D$. 
The results of the experiments are shown at the bottom of Fig.\,\ref{fig_palc}. 
The profiles show the first stage of the PALC work, and the addition of the second stage 
shows a significant improvement in terms of contrast, even though the final contrast is 
limited by various practical factors including the phase-shift in the HEBS apodizer. 
More detail about the results is described in 
\citet{Abe2008}, \citet{Venet2005}.
At present, the PALC is not a highly prioritized candidate for SPICA for various reasons. 
Throughout the design of the SPICA satellite,
it was turned out that the vibrations created by the mechanical
cryo-coolers and the altitude control system are more serious problem
than previously considered.
So a coronagraph which relies on a focal plane mask is not the 
preferred option. 
Furthermore, the success of the laboratory demonstrations of the PIAA/binary-mask hybrid 
coronagraph affords us a high performance 
solution which has now been given priority following the baseline solution provided by 
the binary pupil mask coronagraph. 
Conversely, the result of the MS-PALC experiment is interesting to consider in terms 
of applications for other ground based platforms including current 8\,m class telescopes
(e.g., SUBARU) and future 30m class telescopes (e.g., TMT, EELT, GMT), 
in which a very small occulting mask can be a strong advantage.

\subsection{
PSF subtraction
}
\label{sec_subtraction}

\begin{figure}[!ht]
\begin{center}
\includegraphics*[height=15cm, angle=-90]{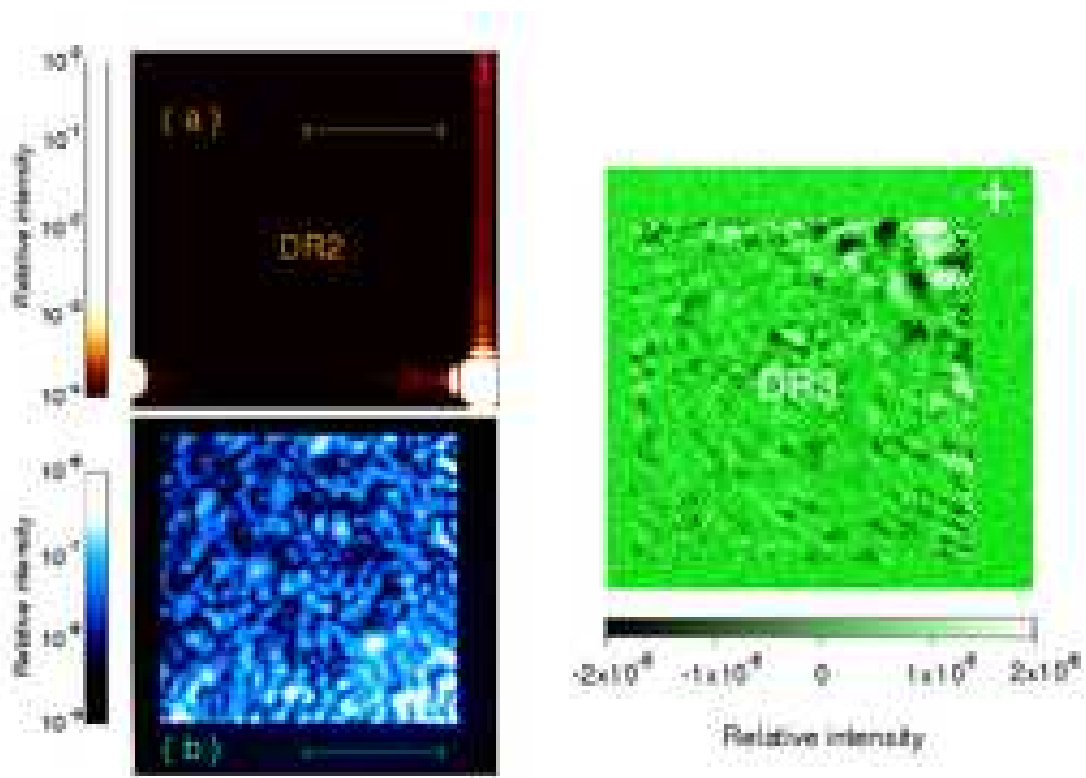}
\end{center}
\caption{
Figures showing the experiment of PSF subtraction of coronagraphic images with the checkerboard Mask-3. 
Top-left: observed coronagraphic images. Panel (a) shows images including the core of the PSF. 
Panel (b) is an image of the dark region obtained with a mask with a square aperture. 
The scale bar shows 10$\lambda/D$. Top-right: an image of the dark region by PSF subtraction. 
Bottom: Profiles of the coronagraphic images obtained by the experiment.
}
\label{fig_subtraction}
\end{figure}

In the case of space telescopes which are perfectly free from air turbulence, 
WFE caused by imperfections in the optics are 
an important limiting factor affecting the contrast of the coronagraph. 
Such WFEs are expected to be quasi-static, 
while they can change over a longer time scale; for example, by thermal deformation of each 
of the elements of the optics. 
PSF subtraction is a useful tool for canceling the stable component of WFE and for achieving 
higher contrast than the basic PSF contrast. 
Following the experiment with the checkerboard Mask-1 and Mask-2, 
we tested the effects of PSF subtraction in our optics,
and the manufacture of a third mask, Mask-3, had been completed for this experiment
as shown in Fig.\,\ref{fig_subtraction}. 
A contrast of 10$^{-10}$ is required in the design of Mask-3, and an optimal 
solution was found. 
The nano-fabrication method used for Mask-3 is basically the same as that used for 
Mask-1 and for Mask-2 as shown in Sec.\,\ref{exp_ckb}.
The configuration of the whole optics was basically the same as that used for the 
demonstrations with Mask-1 and Mask-2, but minor improvements were adopted as shown in 
\citet{Enya2008b}. 
We performed a series of ten, 1 hour exposure-time acquisitions. 
Consecutive exposures were compared by simply subtracting them, after some basic image processing
as used in the case of the experiments with Mask-1 and Mask-2. 
The PSF contrast obtained with Mask-3 is 6.7$\times$10$^{-8}$, which was derived in the same way 
as the data analysis for Mask-1 and Mask-2. The speckle fluctuation over the entire dark 
region for all 10 images is estimated to be $\sim$1.5$\times$10$^{-8}$ rms compared 
to the peak intensity of the PSF core. 
We relate these fluctuations to changes in temperature, 
which could have affected the mechanical stability of the optical bench during the 10 
hour-long observation sessions. 
On the best two images that were continually obtained, 
the intensity fluctuation was 6.8$\times$10$^{-9}$ rms, which is considered to be the data that 
was obtained at the most thermally stable time. 
More detail about this experiments is described in \citet{Enya2008b}.

The most realistic operation for the PSF subtraction with SPICA is image subtraction 
between different stars, in which one star is the target for the study and the other is used 
as a point-source reference. 
It is preferable that the target star and the reference have similar spectrum and luminosity. 
If rolling the telescope were to be possible during the observations as discussed in 
\citet{Trauger2007}, 
in principle any static speckles that originate from aberrations within the whole 
of the telescope system will not move within the field of view, while a planet would rotate.
Therefore the effect of the telescope-roll subtraction will provide the signal of the planet
with cancellation of the static speckles.
In the telescope-roll subtraction, the spectrum and the luminosity of the targeted star 
and the reference star are same, which is an advantage of this method.
As a case study for SPICA, we estimated that a rotation angle of about 8 degrees would be required 
to detect a planet at 6$\lambda/D$ distant from the central star, such that its intensity is not 
reduced by more than 50\% with the telescope-roll subtraction. 
This also puts additional constraints on the minimum distance at which such detection can 
be performed if telescope-roll subtraction is assumed for SPICA,
in which larger roll angle is not realistic because of its thermal design.


\subsection{
Binary mask for the SPICA pupil
}\label{sec_spica_pupil}

A serious issue for the design of the SPICA coronagraph is to find a way of realizing a 
sufficiently small value for IWA. In mid-infrared observations, the spatial resolution of 
the SPICA telescope (as determined by the diffraction limit) is 10 or more times lower 
than optical telescopes of same size. 
Furthermore, the pupil of SPICA is obstructed by a large secondary mirror and its associated 
support structure. 
Large central obstructions seriously affect the IWA values of some types of coronagraph. 
Indeed, the IWA of our checkerboard Mask-1 assuming the obstruction is 7$\lambda/D$,
This value of IWA is more than $\sim$2 times larger than the IWA for Mask-2 (3$\lambda/D$). 
assuming that the pupil is without obstructions (Sec.\,\ref{exp_ckb}). 
An IWA of 7$\lambda/D$ is not satisfactory to perform baseline surveys of nearby stars at a 
distance of 10pc from the Sun. Thus, we needed a counter measure. 

\begin{figure}[!ht]
\begin{center}
\includegraphics*[height=6cm, angle=-90]{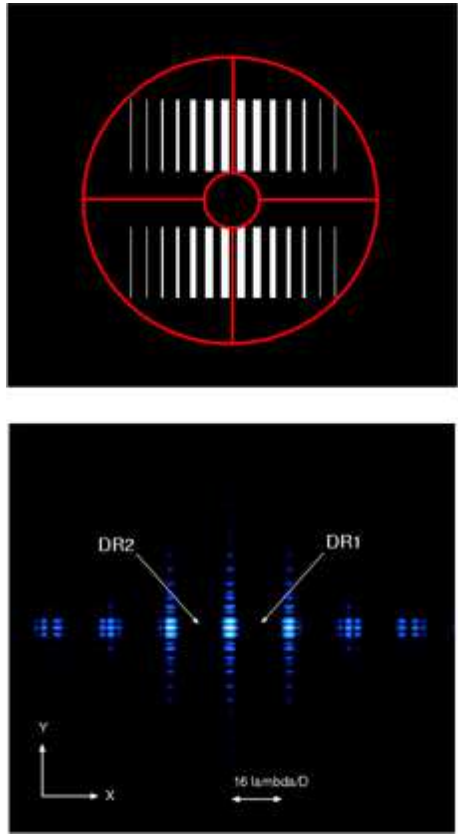}
\end{center}
\caption{
Top: An example of a double bar-code mask for a telescope pupil 
obscured by a secondary mirror and its support structures. 
The transmissivities of the white and black areas are 1 and 0,
respectively. 
The red lines indicate the positions of the obscuring structures. 
Bottom: Simulated PSF using the mask shown in the top panel. 
DR1 and DR2 are the dark regions of the PSF. 
It must be noted that optimization of this sample was performed 
using the LOQO solver presented by \citet{Vanderbei1999}.
}
\label{fig_barcode}
\end{figure}

The top panel in Fig.\,\ref{fig_barcode}
shows an example of our current solution for SPICA, 
which consists of a binary pupil mask, while the bottom panel shows the expected PSF 
derived by simulation. 
This mask is essentially made from a couple of the bar-code masks that were presented in 
\citet{Kasdin2005b}, 
which exhibit coronagraphic power in only 1-dimension. 
It must be noticed taht optimization of this sample was performed using the LOQO solver
presented by \citet{Vanderbei1999}.
It is an interesting point of view to regard this solution as the limit of the asymmetric 
checkerboard mask shown in 
\citet{Tanaka2006a}, \citet{Tanaka2006b}.
In the PSF shown in Fig.\,\ref{fig_barcode}, 
the horizontal direction is coronagraphic. 
The size of the secondary mirror in this solution does not reduce the IWA at all, 
but reduces only the throughput of the mask. 
The central obstruction against coronagraphic power is determined by the width of the 
support structure, which is a lot smaller than the diameter of the secondary mirror. 
As a result, the expected IWA is $\sim$3.3$\lambda/D$, which is as small as the value provided 
by the checker-board pupil without the central obstruction. 

In terms of the PSF of this solution, the diffraction pattern spreads in only two directions, 
i.e. the up and down directions in Fig.\,\ref{fig_barcode}, 
while the diffraction pattern spreads 
in four directions in the case of the checkerboard mask as shown in Fig.\,\ref{fig_ckb_exp}. 
Thanks to this property, the discovery angle around the PSF core becomes larger, 
which is especially beneficial for observations of objects close to the IWA. 
Though the geometry of the mask should be optimized after the final determination of the 
size of the secondary mirror and the width of the support of the SPICA telescope, 
the concept of the double bar-code mask shown above is of general use. 
It is possible to apply the double bar-code mask not only for SPICA, but also for many other 
telescopes in which the pupil is obstructed by a secondary mirror and its supports, 
for example, for SUBARU and other telescopes.

\subsection{
Toward an MIR coronagraph for SPICA
}\label{sec_mirc}

\begin{figure}[!ht]
\begin{center}
\includegraphics*[width=6cm]{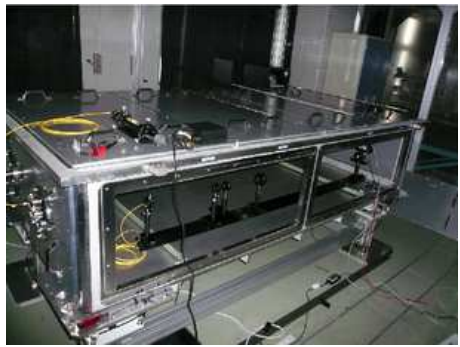}
\end{center}
\caption{
A vacuum chamber for use as a general purpose test-bed for coronagraphic experiments.
}
\label{fig_hoct}
\end{figure}

All of our laboratory experiments that are shown above were performed at room temperature, 
in atmosphere, at visible wavelengths, using a monochromatic laser as a light source. 
Contrarily, the SPICA coronagraph will finally have to be evaluated at cryogenic temperature, 
in vacuum, at infrared wavelengths, using a source with some band-width. 
Thus, we are developing a vacuum chamber for use in coronagraphic 
experiments (Fig.\,\ref{fig_hoct}). 
The most important use of the chamber will be the demonstration of a cryogenic mid-infrared 
coronagraph, while the chamber also has the potential to improve our experiments with 
coronagraphs even at room temperature, owing to the reduction in air turbulence and temperature 
variability. 
 (Indeed, in this conference,  \citet{Haze2008}
present our first results from laboratory 
experiments using a binary-shaped checkerboard mask coronagraph that was fitted inside a 
vacuum chamber in the context of a coronagraphic experiment at visible wavelengths at room 
temperature). 

In our laboratory experiments, small ($\sim$2mm size) coronagraphic masks on glass substrates 
were used to demonstrate the principles that are required to attain the contrast requirements 
for SPICA, i.e. 10$^{-6}$. 
However, to construct a mid-infrared coronagraph we require free-standing 
coronagraphic masks that do not need to be 
positioned on substrates, because free-standing masks are influenced by neither ghosting 
by multi-reflections nor is there any wavelength dependence related to the properties of the 
infrared-transmissive substrates.

\begin{figure}[!ht]
\begin{center}
\includegraphics*[width=14cm]{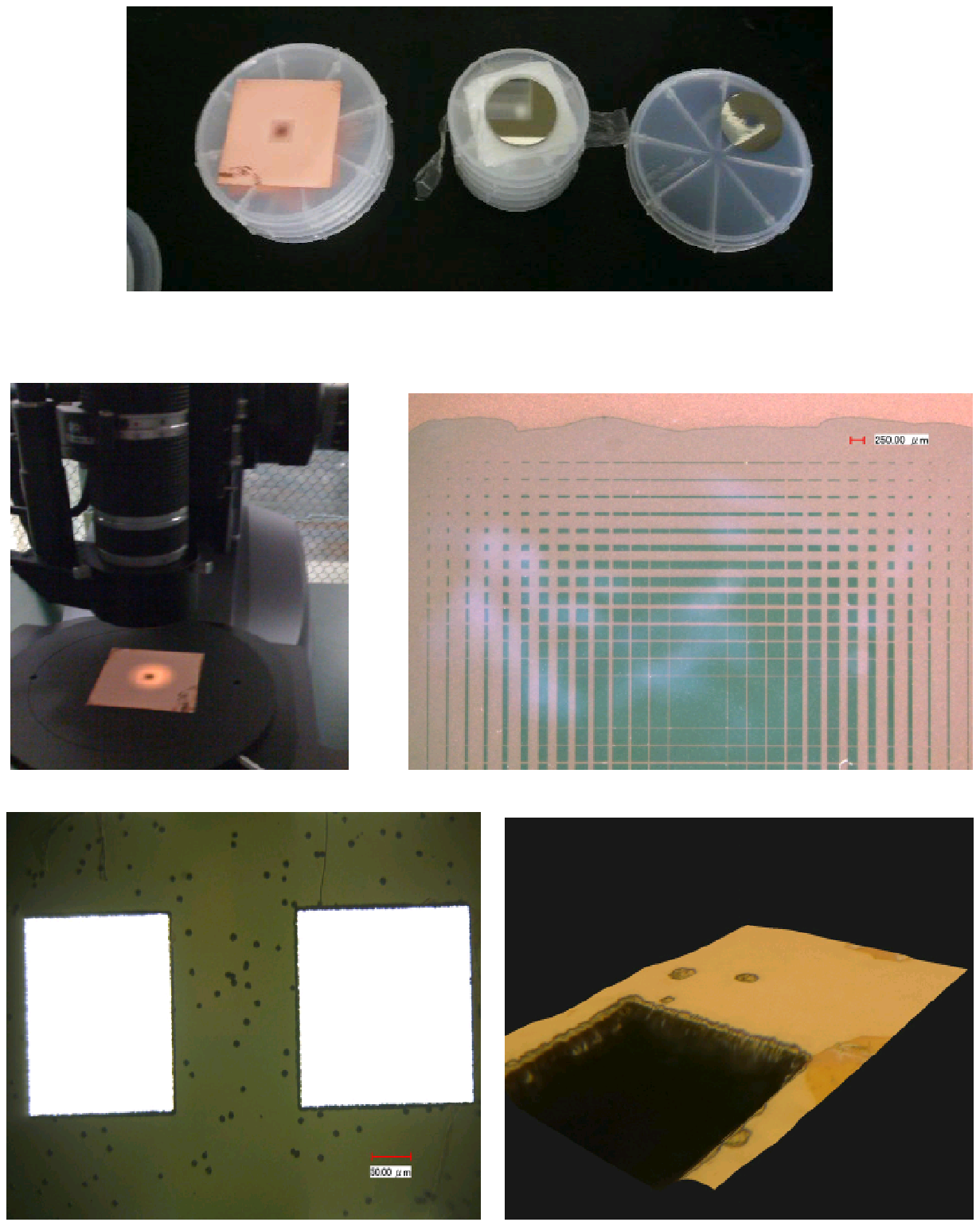}
\end{center}
\caption{
Top: binary-shaped pupil masks produced by trial-and-error 
manufacturing tests using variations of photolithography and etching. 
The left-hand and right-hand devices are free-standing masks, 
while the middle shows masks manufactured on glass substrates to use 
as a reference for comparison. 
Middle-left: a sample of a free-standing mask made of copper set on a
microscope. 
The sizes of whole the device and of the checkerboard pattern 
in the device are 50mm$\times$50mm and 10mm$\times$10mm, respectively.
The thickness of the checkerboard pattern area of this mask is 20$\mu$m, 
while the thickness of the outer area is $\sim$100$\mu$m.. 
Middle-right: a microscope image of the mask taken with reflected
light 
(i.e., light originating from the same side to the imaging side). 
The diffused white pattern is caused by the microscope, 
i.e., it is not intrinsic in the mask. The red scale bar 
indicates 250$\mu$m.
Bottom-left: a microscope image of the same mask taken with 
transmitted light (i.e., light originating from the other side 
to the imaging side). 
The red scale bar shows 50$\mu$m.
Bottom-right: 3-dimensional microscope image of the same mask. 
It was confirmed that there are small pits on the mask, 
but they do not penetrate all of the way through it. 
}
\label{fig_mask}
\end{figure}

We consider that dimensions of $\sim$10mm is a realistic beam size 
in our mid-infrared instrument because of limits 
of the fabrication process to obtain a mask with sufficient accuracy, 
flatness and strength. 
The use of a smaller beam size makes the relative accuracy 
of the mask pattern worse, while the use of a larger beam size makes 
the instrument bigger and heavier, which is quite a serious problem 
that may threaten the decision to let SPICA carry our "optional"
instruments. 
Furthermore, 10mm is comparable in size to the type of MEMS deformable 
mirror with 32$\times$32 channels that can currently be provided 
commercially by BMC. 
Thus, the development of free-standing masks of this size is ongoing, 
and masks manufactured by different methods 
(shown in part in Fig.\,\ref{fig_mask}) are being evaluated and compared.

Many tests are planned for these masks, including microscopic observations of the masks, 
flatness checks by optical 3-dimensional measurements, 
evaluation of coronagraphic performance 
in the visible wavelength, and finally demonstrations of capability in the mid-infrared. 
The development of these masks is now in a phase where they are undergoing manufacturing 
by trial-and-error.

\section{
Active optics
}

\subsection{Cryogenic deformable mirror}

The specification for the image quality of the SPICA telescope requires a 5$\mu$m
diffraction limit, which corresponds to a WFE of 350nm rms, 
and this is not sufficient for a coronagraphic instrument. 
Therefore, we need active optics to compensate for WFE due to the SPICA telescope. 
Such device like a deformable mirror for SPICA  have to work at cryogenic 
temperatures 
(\citet{Dyson2001}; \citet{Mulvihill2003}).
One of the more usual types of deformable mirror is based on the piezoelectric effect; however, 
this effect tends to be smaller at cryogenic temperatures (the piezoelectric 
coefficient at 5K is $\sim$1/10 of that at ambient temperature: see also \citep{Mulvihill2003}). 
Recently, deformable mirrors fabricated using MEMS technology have been developed 
(e.g., \citet{Bifano1997}; \citet{Stewart2007}).
The operation of the MEMS deformable mirror is based on Coulomb forces, 
so does not depend on temperature
like a device using the piezoelectric effect. 
The possibility of realizing large format and compact deformable 
mirrors is another advantage of MEMS technology. 
For example, MEMS deformable mirrors with 32$\times$32 channels are commercially available at BMC. 
However, the stress introduced into MEMS deformable mirrors by 
cooling due to mismatched coefficients of 
thermal expansion (CTE) can cause distortion, and can even destroy the mirrors. 

\begin{figure}[!ht]
\begin{center}
\includegraphics*[height=16cm, angle=-90]{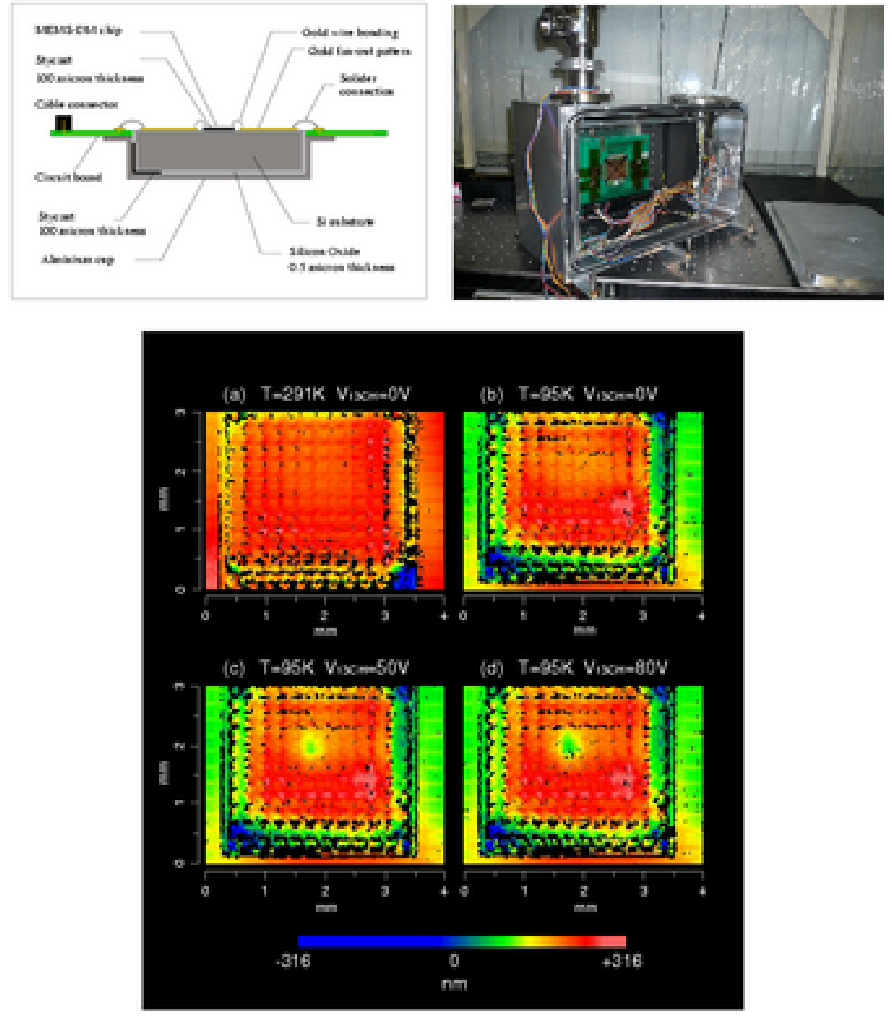}
\end{center}
\caption{
Top-left: schematic view of a prototype deformable mirror unit. 
Top-right: The deformable mirror unit installed in a cryostat. 
Bottom: 3-dimensional surface data obtained by experiments. 
All of the data was acquired through the window of the vacuum cryostat. 
(a), (b), (c), and (d) show the surface without an applied voltage at room temperature, 
without an applied voltage at 95K, with 50V on the 13th channel at 95K, 
and with 80V on the 13th channel at 95K, respectively.
}
\label{fig_dm.eps}
\end{figure}

Hence, we intend to develop and evaluate a prototype cryogenic deformable mirror system 
based on a MEMS device in order to minimize distortion and to prevent 
it from being permanently damaged by thermal stresses introduced 
by cooling as shown in Fig.\,\ref{fig_dm.eps}. 
The silicon substrate is oxidized for electric insulation, 
and then a fan-out pattern of gold is fabricated on the surface. 
A MEMS deformable mirror chip of silicon with 32 channels is bonded to the substrate. 
For the cryogenic tests, we constructed a measurement system consisting of a Fizeau interferometer, 
a cryostat cooled by liquid nitrogen, zooming optics, and electrical drivers
on a optical bench with air suspension in a clean room. 
The surface of the mirror deformed at 95K in response to the application of a voltage, 
and no significant difference was found between the deformation at 95K and that 
at room temperature (Fig.\,\ref{fig_dm.eps}). 
The heat generated was also measured, and
it was suggested that the heat generation is small enough for use 
in a space cryogenic telescope, in which the generated heat 
should be $\sim$1mW or less. 
The properties of the deformable mirror remained unchanged after 5 cycles of vacuum pumping, 
cooling, heating, and venting. We conclude that fabricating 
cryogenic deformable mirrors employing MEMS 
technology is a promising approach.

\subsection{Cryogenic tip-tilt mirror}

\begin{figure}[!ht]
\begin{center}
\includegraphics*[width=8cm]{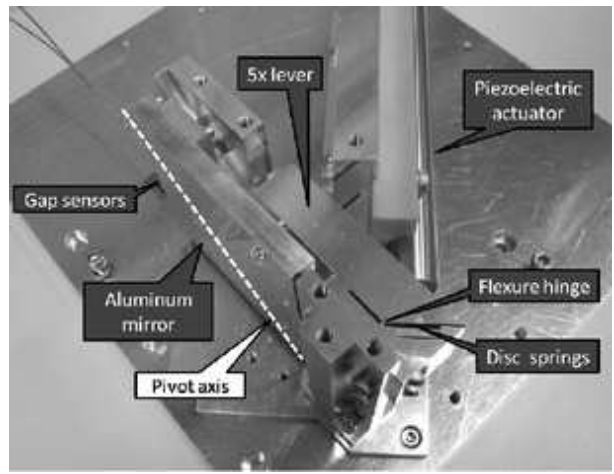}
\end{center}
\caption{
Internal structure of a prototype of a cryogenic tip-tilt mirror
system. 
A 6 cm-square flat mirror is driven by a piezoelectric actuator. 
It should be noted that this was constructed at a laboratory 
pioneered by T. Miyata, S. Sako, T. Nakamura and their colleagues. 
More detail is shown in \citet{Nakamura2008}.
}
\label{fig_chopper}
\end{figure}

The SPICA telescope is influenced by the mechanical vibrations caused by cryo-cooler system
and instruments for altitude control. 
It is expected that such vibrations could correspond to a telescope pointing jitter 
of $\sim$arcsecond, 
which is not acceptable for the coronagraph instrument and the other MIR instruments used in SPICA. 
Therefore, we are developing a cryogenic tip-tilt mirror to compensate for pointing jitter. 
Some prototypes of the cryogenic tip-tilt mirror have been manufactured in collaboration 
with a group for ground-based astronomy at the Institute of Astronomy, University of Tokyo, 
in which a cryogenic chopping mirror has been developed and tested 
in an MIR camera, the MAX\,38 \citep{Miyata2008}. 
Fig.\,\ref{fig_chopper}
shows a prototype for one-axis chopping by a piezoelectric actuator. 
The development of the piezoelectric actuator itself is ongoing 
in order to obtain sufficient stroke in a cryogenic environment 
to counter the stroke-reduction effect of cooling. 
The physical length of the actuators has been enlarged by the addition 
of multi-layers and improvements to the piezoelectric materials 
themselves are being attempted. 
It must be noted that this has been carried out at a laboratory 
pioneered by T. Miyata, S. Sako, T. Nakamura and their colleagues. 
More detail is shown in \citet{Nakamura2008}.

\section{Summary}

SPICA is a proposed JAXA-ESA mission, which was formerly called  H\,II/L2.
SPICA will carry a telescope with a 3.5m diameter monolithic primary mirror, 
and the whole telescope will be cooled to 5K in orbit. 
It is planned that SPICA will be launched in 2017 into the Sun-Earth L2 libration halo orbit 
by an H\,II-A rocket of JAXA for observations 
mainly in the mid and far-infrared. 

The potential of SPICA as a platform for high-contrast observations is discussed in 
comparison with JWST and with ground-based 30m class telescopes, 
which will commence operations from 2010 onwards. 
We suggest that the SPICA mission will be able to provide us with a unique 
and essential opportunity 
for high-contrast observations in the 2010s because of its large telescope aperture, 
its simple pupil shape, capability for infrared observations from space.

We have been performing research and development into a coronagraph for SPICA, 
which is at present regarded as an option for inclusion among the focal plane instruments. 
The strategy for a baseline survey and the specifications of the coronagraph are introduced together, 
in which the primary target of the SPICA coronagraph 
is the direct detection and characterization of Jovian exo-planets 
within a distance of $\sim$10pc from the Sun. 
The main wavelengths for the observations, and the contrast required for 
the SPICA coronagraph instrument are set to be 3.5--27$\mu$m, 
and 10$^{-6}$, respectively. 
Other science case with coronagraphic observation, 
and potential of transit monitoring method for characteization of 
exo-lanets are also presented.

We then summarize the results of our laboratory experiments which are intended to 
demonstrate the principles of the coronagraphs. The experiments were executed in an air 
atmosphere using a visible He-Ne laser as a light source. 
First, we focused on the coronagraph with a binary-shaped pupil mask.
This solution was chosen as a baseline solution for SPICA because of its robust properties 
against telescope pointing-errors and its achromatic performance, 
along with the simplicity of its optics. 
Three binary masks of the checkerboard-type were designed, and masks consisting of aluminum 
film were manufactured with an electron beam on glass substrates using nano-fabrication technology. 
In an experiment involving binary-shaped pupil coronagraphs without active wavefront control, 
a contrast of 6.7$\times$10$^{-8}$ was achieved, as derived from the linear average of the dark 
region and the core of the point spread function. 
On the other hand, a study of PIAA was also started 
in an attempt to achieve higher performance (i.e., smaller IWA and higher throughput). 
The concept of the PIAA/binary-mask hybrid was presented to improve the feasibility of 
manufacturing and the band-width limitations that are caused by diffraction. 
A laboratory experiment was performed using the PIAA/binary-mask hybrid
coronagraph in collaboration at a laboratory developed by O. Guyon and his colleagues. 
This experiment employed active wavefront control, 
and a contrast of 6.5$\times$10$^{-7}$ was achieved. 
These contrasts surpass the requirements for the SPICA coronagraph, which are set at 10$^{-6}$.
A PALC has also been studied. 
Apodizer masks made of HEBS glass have been manufactured to demonstrate a MS-PALC,
in which it was confirmed that the inclusion of the second stage significantly 
improved the performance. 

We also present recent progress in the cryogenic active optics for SPICA. 
Prototype devices of cryogenic 32-channel deformable mirror
fabricated by MEMS techniques were developed, 
and the first cryogenic demonstration of surface deformation was performed with 
liquid nitrogen cooling. 
Experiments involving piezoelectric actuators for cryogenic tip-tilt mirrors are also ongoing. 

In our future work, the SPICA coronagraph will finally have to be evaluated at cryogenic 
temperatures, in vacuum, at infrared wavelengths, using a source with some band-width. 
Thus, we are developing a vacuum chamber for just such coronagraphic experiments. 
The development of free-standing masks of $\sim$10mm in size is ongoing, and masks 
manufactured by different methods are being evaluated and compared. 
A design using a double bar-code mask is presented, 
which exhibits coronagraphic power in only one direction,
as an example of our current baseline 
solution for the actual pupil of SPICA which includes some obstructions.

This paper summarizes our studies accomplished until the kick-off of the pre-project phase of SPICA. 
We consider that SPICA will provide a unique and essential 
opportunity for the direct observation of exo-planets, and studies
for critical basic technologies are set to succeed. 
Therefore, we propose to develop a mid-infrared coronagraph instrument for SPICA and to 
perform the direct observation of exo-planets with SPICA.

\section{Acknowledgments}

We are grateful to T. Wakayama, T. Sato, and N. Nakagiri in the Nanotechnology Research 
Institute of AIST. 
We would like to give thanks for kind support by N. Okada, T. Nishino, T. Fukuda, K. Kaneko 
at the machine shop, and K. Mitsui, M. Yokota at the optics shop in the Advanced Technology 
Center of the National Astronomical Observatory of Japan. 
We also give thanks for the effort supplied by T. Yamada, M. Hayashi and all related 
personnel to realize the collaboration with the SUBARU Observatory. 
We thank with respect the pioneers of advanced shaped pupil mask 
coronagraph design, especially  R. Vanderbei, J. Kasdin, and
R. Belikov. 
We are grateful for fruitful comments from M. Ferlet, B. Swinyard, 
T. Matsuo, M. Fukagawa, Y. Itoh, T. Yamashita, N. Narita, J. Schneider, 
A. Boccaletti, A. Burrows, D. Spiegel, D. Deming, and G. Tinetti. 
We also give thanks for the deep understanding of this work  of the
referees and the productive support of the editors. 
The works shown in this paper are supported by a grant from JAXA, the Japan Society for 
the Promotion of Science, and the Ministry of Education, Culture, Sports, Science 
and Technology of Japan. 
We would like to express special gratitude to S. Tanaka,
even after the change of his field.
Lastly, the author is deeply grateful to L. Abe, O. Guyon and 
all of the SPICA working group.

\clearpage

\newpage

\section{Tables}

\begin{table}[!ht]
\caption{SPICA Mission specification}
\begin{tabular}{ll}
\hline
\hline
Item        & Specification  \\
\hline
Telescope aperture & 3.5m diameter \\
Telescope temperature  &  5K (during observation), 300K (launch) \\
Core wavelength of observation  & 5--200$\mu$m     \\
Wavefront quality by telescope  & 5$\mu$m diffraction limit  \\
Cryogenics       &  Mechanical cryo-coolers \& radiation cooling  \\
Launch year      &  2017       \\
Launch vehicle    &  H\,II-A \\
Orbit            &  Sun-Earth L2 Halo       \\
Mission life     &  $\geq$5 years \\
\hline
\end{tabular}
\label{table1}
\end{table}

\vspace{20mm}

\begin{table}[!ht]
\caption{Specification of the SPICA coronagraph}
\begin{tabular}{ll}
\hline
\hline
Item        & Specification  \\
\hline
Sensitive wavelength  &  3.5--27$\mu$m (shorter wavelength is optional) \\
Observation mode & Imaging, spectroscopy       \\
Coronagraphic method   &  Binary-shaped pupil mask (baseline)  \\
                       &  PIAA/binary-mask hybrid (optional)          \\
Contrast   & 10$^{-6}$           \\
Inner working angle (IWA)  & $\sim$3.3$\lambda/D^*$ (binary mask) \\
                          & $\sim$1.5$\lambda/D$ (PIAA/binary-mask hybrid option) \\
Outer working angle (IWA)  & $\geq$16$\lambda/D$       \\
Detector                  & $\sim$1k$\times$1k format Si:As array  (InSb detector is optional)\\
Field of view    &  $\sim$\,1'$\times$1'          \\
Spectral resolution         & $\sim$20--200           \\
\hline
\end{tabular}
\label{table2}
* $\lambda$ and $D$ is the observation wavelength and the diameter 
of the telescope aperture(3.5m), respectively. 
\end{table}

\newpage
\section{Figures}


\newpage


\end{document}